\documentclass[aps,pra,twocolumn,amsmath,amssymb,superscriptaddress,showpacs]{revtex4}
\usepackage{graphics}
\usepackage{graphicx}
\usepackage{bbm}
\usepackage{amsfonts}
\usepackage{amssymb}
\usepackage{epsfig}
\usepackage{subfigure}
\usepackage{float}
\usepackage{CJK}

\begin{document}

\title{Lifshitz phase transitions in one-dimensional Gamma model}
\author {Zi-An Liu}
\affiliation{College of Science, Nanjing University of Aeronautics and Astronautics, Nanjing 211106, China }
\affiliation{School of Physical Science and Technology, Soochow University, Suzhou, Jiangsu 215006, China}
\author {Tian-Cheng Yi}
\affiliation{College of Science, Nanjing University of Aeronautics and Astronautics, Nanjing 211106, China }
\affiliation{School of Physical Science and Technology, Soochow University, Suzhou, Jiangsu 215006, China}
\author {Jin-Hua Sun}
\affiliation{Department of Physics, Ningbo University, Ningbo 315211, China}
\author {Yu-Li Dong}
\affiliation{School of Physical Science and Technology, Soochow University, Suzhou, Jiangsu 215006, China}
\author {Wen-Long You}
\email{wlyou@nuaa.edu.cn}
\affiliation{College of Science, Nanjing University of Aeronautics and Astronautics, Nanjing 211106, China }
\affiliation{School of Physical Science and Technology, Soochow University, Suzhou, Jiangsu 215006, China}

\begin{abstract}
In this paper, we study quantum phase transitions and magnetic properties of a one-dimensional spin-1/2 Gamma model,
which describes the off-diagonal exchange interactions between edge-shared octahedra with strong spin-orbit couplings along the sawtooth chain. The competing exchange interactions between the nearest neighbors and the second neighbors stabilize semimetallic ground state in terms of spinless fermions, and give rise to a rich phase diagram, which consists of three gapless phases. We find distinct phases are characterized by the number of Weyl nodes in the momentum space, and such changes in the topology of the Fermi surface without symmetry breaking produce a variety of Lifshitz transitions, in which
the Weyl nodes situating at $k=\pi$ interchange from type I to type II. A coexistence of type-I and type-II
Weyl nodes is found in phase II. The information measures including concurrence, entanglement entropy and relative entropy can effectively signal the second-order
transitions. The results indicate that the Gamma model can act as an exactly solvable model to describe Lifshitz phase transitions in correlated electron systems.
\end{abstract}
\maketitle

\section{Introduction}\label{introduction}
The strong quantum fluctuations in low-dimensional magnets often give rise to fascinating quantum phenomena, which have been intensively investigated. Usually the underlying exotic physics in these complex systems are hard to identify. In this context, a few exactly solvable models provide particular enlightenment for the analysis of enigmatic quantum phase transitions (QPTs) in correlated electron systems. Experimental realizations are, for example, the Ising-type materials LiHoF$_4$ \cite{Bitko96}, CoNb$_2$O$_6$ \cite{Coldea10}, XY-type materials Cs$_2$CoCl$_4$ \cite{Kenzelmann02,Breunig13}, and Heisenberg-type magnetic insulator copper pyrazine dinitrate Cu(C$_4$H$_4$N$_2$)(NO$_3$)$_2$ (denoted by CuPzN for short) ~\cite{Breunig17}, in which the spins at each transition metal ions are structurally arranged along the crystallographic $a$ axis to form a spin-1/2 chain. Recently enormous research efforts have been devoted to the celebrated Kitaev model~\cite{Kit06}, which has become one of prototype models that support the spin-liquid ground state and the associated non-Abelian quasiparticles.
The Kitaev spin liquid is exactly realized in the Kitaev honeycomb lattice, which is characterized by the Ising-type interactions ($\propto$ $S_i^a$$S_j^a$) between nearest-neighbor spins with different easy-axis directions depending on three types of bonds ($a$ = $x$, $y$, $z$) on the honeycomb lattice.  The Kitaev model was initially believed to be tractable mathematically, and then become physically realistic since Jackeli and Khaliullin~\cite{Jackeli09} demonstrated that the bond-directional interactions could be realized in Mott insulators with strong spin-orbit coupling.

With the rapid progress in materials synthesis and heterostructure design, the growing interest in orbital degrees
of freedom and spin-orbital coupling for strongly correlated electrons in transition metal oxides was amplified by the exotic quantum state discovered.  For instance, the interfacial Dzyaloshinskii-Moriya interaction (DMI) in graphene/ferromagnet heterostructures was observed~\cite{Chaurasiya19}.
In this regard, various effective models for one-dimensional (1D) and one-dimensional-analogue architectures are devised. In Ref. [\onlinecite{You2014}], one of the authors proposed a 1D compass model on the sawtooth-chain ferromagnetic transition metal oxides with active $e_g$ orbitals, which shared similar characteristic feature of bond-dependent Ising-like interactions with the Kitaev model.
Such one-dimensional-analogue nanostructures constitute flexible
platforms for designing simplified models, which provide a remarkable description of the characteristic properties of low-dimensional frustrated systems. Studies of spin liquid state have been hampered by the lack of a simple solvable model that can capture features of exotic topological state of matter.\\
\indent In this paper, we consider an effective Hamiltonian by taking into account the off-diagonal exchange interactions for edge-shared octahedra structures with strong spin-orbit couplings along a snake-like chain. Because of the sawtooth architecture, the system is not only affected by the nearest-neighbor interactions, but also by the second-nearest-neighbor interactions. We study the QPTs in 1D system including both the off-diagonal exchange interactions and the three-site interactions. The merit of this model is that it depicts
a quantum spin chain with broken reflection symmetry, which is not only physically realistic, but also tractable mathematically.\\
\indent The organization of the rest of the paper is as follows. In Sec. \ref{model}, the Hamiltonian model is given and an analytical expression of the ground-state energy is obtained. We find the low-energy dispersions around $k$=$\pi$ in momentum space play a crucial role in determining the topology of the Fermi surface. The phase diagram is then established in the $\Gamma_2$-$\eta$ plane. In Sec. \ref{correlation function}, we show that the correlation functions can be analytically obtained. In Sec. \ref{Quantum entanglement}, based on the exact solution of correlations, information measures including concurrence, von Neumann entropy and relative entropy are adopted to locate the quantum critical points, which implies that these information measures act as universal order parameters for Lifshitz phase transitions in many-body lattice systems.

\section{Model and phase diagram}\label{model}
The Hamiltonian of the Gamma model is given by
\begin{eqnarray}
{\cal H}&=&\sum_{j=1}^{N}[  \Gamma_1(\sigma_j^x \sigma_{j+1}^y+\alpha \sigma_j^y \sigma_{j+1}^x)
     \nonumber \\
    &+& \Gamma_2(\sigma_{j-1}^x \sigma_j^z \sigma_{j+1}^y+\beta \sigma_{j-1}^y \sigma_j^z \sigma_{j+1}^x)],
    \label{Hamiltonian1}
\end{eqnarray}
where $\sigma_j^a$ ($a$= $x$, $y$, $z$) are Pauli matrices on the $j$-th site of an $N$-site system.
$\Gamma_1$ and $\Gamma_2$ denote the strength of nearest-neighbor and next-nearest-neighbor off-diagonal exchange interaction,
$\alpha$ and $\beta$ represent the relative coefficients between different terms of off-diagonal exchange couplings. For later convenience, we define $\eta=\alpha/(2\beta)$. Throughout the paper, we take $\Gamma_1$=1 as the energy unit.
For $\alpha=1$, Eq. (\ref{Hamiltonian1}) becomes the symmetric off-diagonal exchange interactions. It is found that the off-diagonal exchange interactions play a significant role in stabilizing the quantum spin liquid state~\cite{Takikawa19,Rau2014} and ordered phases~\cite{Gordon2019,Yang20,Yang202}. Such form of exchange can also emerge from the truncated dipolar exchange~\cite{Gardner10,Rynbach10}.
Furthermore, for $\alpha=-1$, the antisymmetric form reduces to the DMI, which was first proposed by Dzyaloshinsky and Moriya~\cite{Dzyaloshinsky1958,Moriya1960} and had attracted continued interest~\cite{yi2019,SatyakiKar2018,You142,Qiu16,WuQC17,Huia2017,Liang2017}. The DMI has been proved to be a key factor in explaining the magnetic properties in LiMnPO$_4$~\cite{Ellen2019}, Ni$_3$V$_2$O$_8$~\cite{Kenzelmann2006},  MnSi~\cite{Dhital2017,Shanavas2016} and CoFeB~\cite{Soucaille2016}, etc.
Note that the emergence of three-site interactions has been discussed
in the context of nonequilibrium dynamics as an energy current~\cite{Zotos1997,Antal1997}. We derive the XZY$-$YZX-term from the first two terms of Eq. (\ref{Hamiltonian1}) in the Appendix \ref{energycurrent}. The three-site interactions compete with the nearest-neighbor interactions and generally induce the phase transitions. For instance, the XZY$-$YZX-type three-site interactions in the anisotropic XY spin chain gives rise to gapless phases~\cite{Tong2015,Tong2016,You2016}.

In this paper, we consider even $N$ and impose periodic boundary conditions (PBCs) with $\sigma_{N+1}^a=\sigma_{1}^a$. For convenience, in terms of the raising and lowering operators $\sigma_j^{\pm}$= $(\sigma_j^x$ $\pm$ $i \sigma_j^y)/2$,
the Jordan-Wigner transformation converts the spin operators into spinless fermion operators by the following relations:
\begin{eqnarray}
&&\sigma_j^+=e^{i\pi \sum_{n=1}^{j-1}c_n^\dagger c_n}c_j, \sigma_j^-=e^{-i\pi \sum_{n=1}^{j-1}c_n^\dagger c_n}c_j^\dagger, \nonumber \\
 &&\sigma_j^z= e^{\pm i\pi c_j^\dagger c_j},
\label{JW}
\end{eqnarray}
where $c_{j}$ and $c_{j}^{\dagger }$ are annihilation and creation
operators of spinless fermions at site $j$ which obey the standard
anticommutation relations, i.e., $\{c_{i},c_{j}\}=0$ and
$\{c_{i}^{\dagger}$,$c_{j}\}$=$\delta_{ij}$.
Thus, the Hamiltonian (\ref{Hamiltonian1}) can be rewritten as a quadratic form of the creation 
and annihilation operators of spinless fermions:
\begin{eqnarray}
{\cal H}&=& \sum_{j=1}^{N}i\Gamma_1 \left[(c_{j}^\dagger-c_{j})(c_{j+1}^\dagger- c_{j+1})\right.\nonumber \\
&+& \left. \alpha(c_{j}^\dagger+c_{j})(c_{j+1}^\dagger+ c_{j+1})\right]\nonumber \\
&+&i\Gamma_2  \left[(c_{j-1}^\dagger-c_{j-1})(c_{j+1}^\dagger- c_{j+1})\right.\nonumber \\
&+& \left. \beta(c_{j-1}^\dagger+c_{j-1})(c_{j+1}^\dagger+ c_{j+1})\right].
\label{fer}
\end{eqnarray}
Note that the boundary term in Eq.(\ref{fer}) has an extra phase factor $c_{N+1}$= $ c_1 (-1)^{(N_p+1)}$ with  the total fermion number $N_p$=$\sum_{j=1}^N c_{j}^\dagger c_{j}$. Such subtle boundary effect leads to either PBC
or antiperiodic boundary condition (APBC) for the spinless fermion chain~\cite{Lieb61,Cabrera87,Wu20}. The boundary contribution becomes negligible after the thermodynamical limit has been taken owing to the $1/N$ correction.
To diagonalize the Hamiltonian in Eq. (\ref{fer}) in the APBC channel with even fermion-number parity, a Fourier transformation $c_{j}$=$\frac{1}{\sqrt{N}}\sum_{k}e^{-ik j}c_{k}$ is introduced
with the discrete momenta given as follows:
\begin{eqnarray}
k=\frac{n\pi}{ N }, \quad n= -(N-1), -(N-3),
\ldots, N-1.
\end{eqnarray}
The Hamiltonian can be written in the following form:
\begin{eqnarray}
\!{\cal H}\!&=&\!\sum_{k} 2\Gamma_1(\alpha\!-\!1)\sin{k}   c_{k}^\dagger c_{k}
\!+\! \Gamma_1(\!\alpha\!+1\!)\sin{k}  (c_{-k}c_{k}\!+\!c_{k}^\dagger c_{-k}^\dagger)\nonumber\\
\! &+&\! 2\Gamma_2(\!\beta\!-\!1)\sin{2k}c_{k}^\dagger c_{k}
\!+\Gamma_2(\!\beta\!+\!1)\sin{2 k}(c_{-k}c_{k}\!+\!c_{k}^\dagger c_{-k}^\dagger). \quad \;
\end{eqnarray}
Then we write
it in a symmetrized matrix form with respect to the
$k\leftrightarrow -k$ transformation within the Bogoliubov-de Gennes (BdG) representation,
\begin{eqnarray}
{\cal H} &=&  \sum_{k}\,
\Lambda_k^{\dagger}\,\hat{M}_k^{}\,\Lambda_k^{},
\label{FT2}
\end{eqnarray}
where
\begin{eqnarray}
\hat{M}_k=\left(\begin{array}{cc}
A_k &  B_k  \\
B_k^*    &  -A_{-k}   \\
\end{array}\right),\nonumber \\
\label{Mk}
\end{eqnarray}
with $A_k$=$\Gamma_1(\alpha-1)\sin{k}$+$\Gamma_2(\beta-1)\sin{2k}$,
$B_k$=$\Gamma_1(\alpha+1)\sin{k}$+$\Gamma_2(\beta+1)\sin{2k}$ and $\Lambda_k^{\dagger}=(c_k^{\dagger},c_{-k})$. Note that $A_{-k}$=$-A_k$ and $B_{-k}$=$-B_k$. Finally, it can be diagonalized by using the Bogoliubov transformation. In this way, the ground-state energy can be obtained and the Hamiltonian can be transformed into the diagonal form
\begin{eqnarray}
{\cal H}=\sum_k \varepsilon_{k}(b_{k}^\dagger b_{k}-\frac{1}{2}),
\label{Hamiltonian2}
\end{eqnarray}
where the energy spectrum is given by
\begin{eqnarray}
\varepsilon_{k}&=&2(A_k + \vert B_k \vert)\nonumber \\
&=&2 \big[\Gamma_1(\alpha-1)\sin{k}+\Gamma_2(\beta-1)\sin{2k} \nonumber\\
 &+& 2 \vert \Gamma_1(\alpha+1)\sin{k}+\Gamma_2(\beta+1)\sin{2k} \vert\big].
\label{varepsilon}
\end{eqnarray}
\begin{figure}[tb!]
\centering
 \includegraphics[width=6cm]{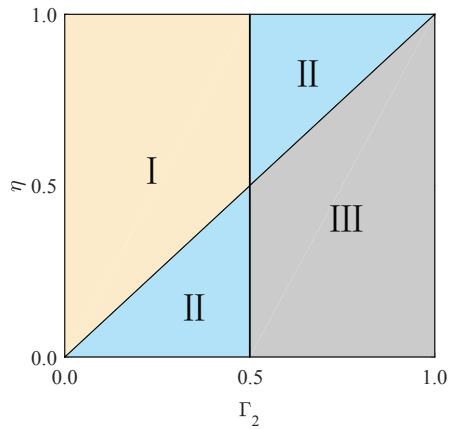}
 \caption{The ground-state phase diagram of the Gamma model with three-site interactions. The critical lines correspond to $(\Gamma_{2})$$_{c,1}$= 0.5  and $(\Gamma_{2})$$_{c,2}$=$\eta$.
 }
\label{phasediagram}
\end{figure}
With the excitation energy at hand, we have precisely determined the ground-state phase diagram of the generalized Gamma model with respect to $\eta$ and  $\Gamma_2$ in Fig.\ref{phasediagram}.
One finds $\varepsilon_{0}$=$\varepsilon_{\pi}$=0, which implies the ground state is always gapless. For the low-energy spectrum around the Fermi point $k_F=0$ with $\vert k\vert\ll 1$, we have
\begin{eqnarray}
\varepsilon_{k}&\approx&  \big[2\Gamma_1(\alpha-1) +4\Gamma_2(\beta-1)\big] k \nonumber \\
&+& \big[ 2 \Gamma_1(\alpha+1)+4\Gamma_2(\beta+1)\big]\vert k \vert,
\label{varepsilon2}
\end{eqnarray}
which implies the bands near the band crossing point disperse linearly. The anisotropy is quite similar to a tilted Weyl cone in the context of Weyl semimetals (WSMs).
The velocity of a left going spinless fermion is $v^{I}_l$= $-4\Gamma_1- 8\Gamma_2 $ and that of a right going spinless fermion is $v^{I}_r$= $4\Gamma_1 \alpha + 8\Gamma_2\beta$.

\begin{figure}[ht]
\includegraphics[width=8cm]{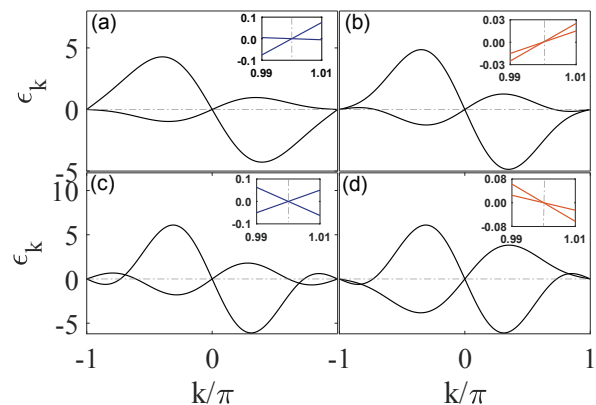}
\caption{The energy spectrum for four representative
sets of $\{\eta, \Gamma_2\}$ parameters which correspond to
different regions as depicted in Fig.\ref{phasediagram}.
(a) $\eta$=0.25, $\Gamma_2$=0.2;                                      (b) $\eta$=0.25, $\Gamma_2$=0.4;                                       (c) $\eta$=0.25, $\Gamma_2$=0.75 and                                     (d) $\eta$=1, $\Gamma_2$=0.75.
Insets in (a),(c) illustrate the type-I dispersions
 around $k=\pi$ and insets in (b),(d) show type-II dispersions around $k=\pi$.}
\label{energyspectrum}
\end{figure}

On the other hand, for the low-energy excitation around the Weyl node  $k_F=\pi$ with $\vert\delta_k \vert\ll 1$, in which $\delta_k= k-\pi$,
\begin{eqnarray}
\varepsilon_{k}&\approx&  \big[4\Gamma_2(\beta-1)-2\Gamma_1(\alpha-1) \big]  \delta_k  \nonumber \\ &+& \vert  4 \Gamma_2(\beta+1)-2\Gamma_1(\alpha+1) \vert \vert  \delta_k \vert.
\label{varepsilon3}
\end{eqnarray}
In phase I, the velocity of a left going spinless fermions is $v^{I}_l$= $-4\Gamma_1 \alpha + 8\Gamma_2\beta$ and that of a right going spinless fermions is $v^{I}_r$= $ 4\Gamma_1- 8\Gamma_2$. The dispersion around $\pi$ shows an up-right tapered shape in the energy-momentum space, resembling a conventional type-I Weyl point [see Fig. \ref{energyspectrum}(a)].
With increasing $\Gamma_2$, one of two branches will become dispersionless at either $(\Gamma_{2})$$_{c,1}$=0.5 or $(\Gamma_{2})$$_{c,2}$=$\eta$. After surpassing the critical point, the system enters phase II with 3 Fermi points. In contrast to phase I, the Weyl cone is tipped over, in which the two crossing bands have the same sign of their slopes along certain directions in the
$k$-space, forming a type-II Weyl point.  Thus a change in
the topology of the Fermi surface without symmetry breaking from the type-I Weyl point in Fig. \ref{energyspectrum}(a) to the type-II Weyl point in Figs. \ref{energyspectrum} (b) and (d) represents a Lifshitz transition. Upon further increasing $\Gamma_2$,  the system is driven into the phase III consisting of 4 Fermi points through another Lifshitz phase transition, in which one of branch becomes again dispersionless. One can find in phase III the two crossing bands at either $k=0$ or $\pi$ always have slopes with opposite signs in $k$-space [see Fig. \ref{energyspectrum} (c)].

The ground state corresponds to the configuration in which all states with $\varepsilon_{k}< 0$ ($\in \Omega_k^-$) are filled and  states with $\varepsilon_{k} > 0$ ($\in \Omega_k^+$) are empty.  Therefore, the ground-state energy density is given by
\begin{eqnarray}
e_0&=&-\frac{1}{2N} \sum_{k\in\Omega_k^+} \varepsilon_{k}+ \frac{1}{2N}\sum_{k\in\Omega_k^-} \varepsilon_{k}.
\end{eqnarray}
In the thermodynamic limit $N$$\to$$\infty$, one can obtain
\begin{widetext}
\begin{eqnarray}
\label{varepsilon4}
e_0&=&  -\frac{1}{ \pi}  \int_{-\pi}^\pi (\vert \Gamma_1 \alpha \sin{k}+\Gamma_2 \beta \sin{2k} \vert
+ \vert \Gamma_1  \sin{k}+\Gamma_2 \sin{2k} \vert)dk \nonumber \\
&=& -\frac{\vert \Gamma_1 \alpha  \vert }{\pi}\left[\frac{1+4\bar{\Gamma}_2^2}{\bar{\Gamma}_2}\Theta(-0.5-\bar{\Gamma}_2)+4\Theta(0.5-\vert \bar{\Gamma}_2\vert) +\frac{1+4\bar{\Gamma}_2^2}{\bar{\Gamma}_2}\Theta( \bar{\Gamma}_2-0.5)\right]\nonumber \\
&-&\frac{\vert \Gamma_1 \vert }{\pi}\left[\frac{1+4\tilde{\Gamma}_2^2}{\tilde{\Gamma}_2}\Theta(-0.5-\tilde{\Gamma}_2)+4\Theta(0.5-\vert \tilde{\Gamma}_2\vert) +\frac{1+4\tilde{\Gamma}_2^2}{\tilde{\Gamma}_2}\Theta( \tilde{\Gamma}_2-0.5)\right],
\end{eqnarray}
\end{widetext}
where $\tilde{\Gamma}_2$=$\Gamma_2/\Gamma_1$, $\bar{\Gamma}_2$=$\Gamma_2 \beta/\Gamma_1 \alpha $ and $\Theta(\cdot)$ is the Heaviside step function, whose value is zero for negative arguments and one for positive arguments.

\begin{figure}[tb!]
\includegraphics[width=8cm]{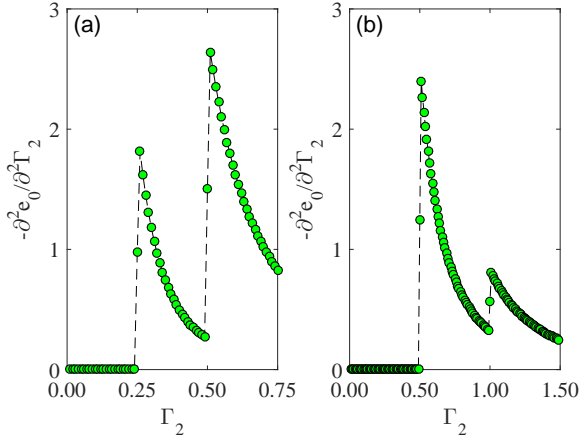}
\caption{The second-order derivative of energy density $e_0$ as a function of $\Gamma_2$ for (a) $\eta=0.25$ and
 (b) $\eta=1$.}
\label{energyderivative}
\end{figure}

\begin{figure}[tb!]
\centering
\includegraphics[width=8cm]{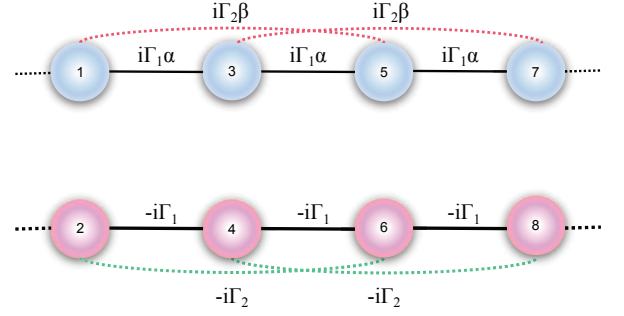}
\caption{Auxiliary snake chain representation for the Hamiltonian in Eq. (\ref{Hamiltonian1}) in terms of Majorana fermions.
}
\label{syt}
\end{figure}

\emph{Symmetry analysis.}
It is easy to find Eq. (\ref{varepsilon4}) remains invariant for $\Gamma_2$ $\to$ $-\Gamma_2$ and $k$ $\to$ $\pi-k$. Therefore, $e_0(\Gamma_2,\alpha,\beta)$=$e_0(-\Gamma_2,\alpha,\beta)$=$e_0(\Gamma_2,-\alpha,\beta)$=$e_0(\Gamma_2,\alpha,-\beta)$. Without any loss of generality, we set $\alpha\ge 0$, $\beta\ge 0$ and $\Gamma_2\ge0$. One can easily observe in Fig. \ref{energyderivative} that the second-order phase transitions at both $(\Gamma_{2})$$_{c,1}$=0.5 and $(\Gamma_{2})$$_{c,2}$=$\eta$ are characterized by a discontinuity in the second derivatives of the energy density.
In fact, two QPTs are triggered by different types of competitions due to a hidden symmetry behind the model ${\cal H}$, which can be unveiled by exploring the corresponding Majorana Hamiltonian. %
Regarding the Majorana modes $\gamma_{2j-1}$=$c_j^\dagger+c_j$, $\gamma_{2j}$=$i(c_j-c_j^\dagger)$, Eq. (\ref{fer}) with open boundary conditions can be rewritten as
\begin{eqnarray}
{\cal H}&&=\sum_{j=1}^{N-1}(-i\Gamma_1\gamma_{2j}\gamma_{2j+2}+i\Gamma_1\alpha\gamma_{2j-1}\gamma_{2j+1}\nonumber
\\&&-i\Gamma_2\gamma_{2j-2}\gamma_{2j+2}+i\Gamma_2\beta\gamma_{2j-3}\gamma_{2j+1}).
\label{Majorana}
\end{eqnarray}
A schematic diagram of Majorana modes is shown in Fig.\ref{syt}, in which two Majorana chains are decoupled in the absence of the magnetic field. Along the odd (even) chain, the interactions between nearest-neighbor sites are $-i\Gamma_1\alpha$ ($i\Gamma_1$) and the interactions between the next-nearest-neighbor are $i\Gamma_2\beta$ ($-i\Gamma_2$). Hence, the Hamiltonian in Eq.(\ref{Hamiltonian1}) can be rewritten as a sum of two commuting parts, i.e., ${\cal H}$= ${\cal H}_{\rm o}$+ ${\cal H}_{\rm e}$, where
\begin{eqnarray}
 {\cal H}_{\rm o}&=&\sum_{j=1}^{N} \Gamma_1 \alpha \sigma_j^y \sigma_{j+1}^x +  \Gamma_2 \beta \sigma_{j-1}^y \sigma_j^z \sigma_{j+1}^x ,\label{oddchain}\\
{\cal H}_{\rm e}&=&\sum_{j=1}^{N} \Gamma_1 \sigma_j^x \sigma_{j+1}^y +
     \Gamma_2 \sigma_{j-1}^x \sigma_j^z \sigma_{j+1}^y.
      \label{evenchain}
\end{eqnarray}
We are aware that the reflection symmetries in Eq. (\ref{oddchain}) and (\ref{evenchain}) are both broken. It becomes evident that two critical points $(\Gamma_{2})$$_{c,1}$=0.5 and $(\Gamma_{2})$$_{c,2}$=$\eta$ origin from two split Majorana chains. For $\alpha=0$ and $\beta$=0,  ${\cal H}_{\rm o}$ turns out to vanish and one can notice the ground state is $2^{N/2-1}$-fold degenerate with PBCs and $2^{N/2}$-fold degenerate with open boundary conditions. The macroscopic degeneracy can be interpreted either by the intermediate symmetry operators~\cite{You08,You14,Nussinov14} or by the free Majorana modes~\cite{Wu19}.  Once a finite local transverse magnetic field $g_l$ $(1\le l\le N)$  is turned on, an $l$th tooth will hybridize the even Majorana chain with a Majorana mode in the odd Majorana chain since $g_l$ $\sigma_l^z $= $ig_l\gamma_{2l-1}\gamma_{2l}$.
 It is intriguing to involve more finite external fields, but in this paper we will restrict the discussion to the zero field limit.

\section{Correlation function}
\label{correlation function}
\begin{figure}[tb!]
\includegraphics[width=7cm]{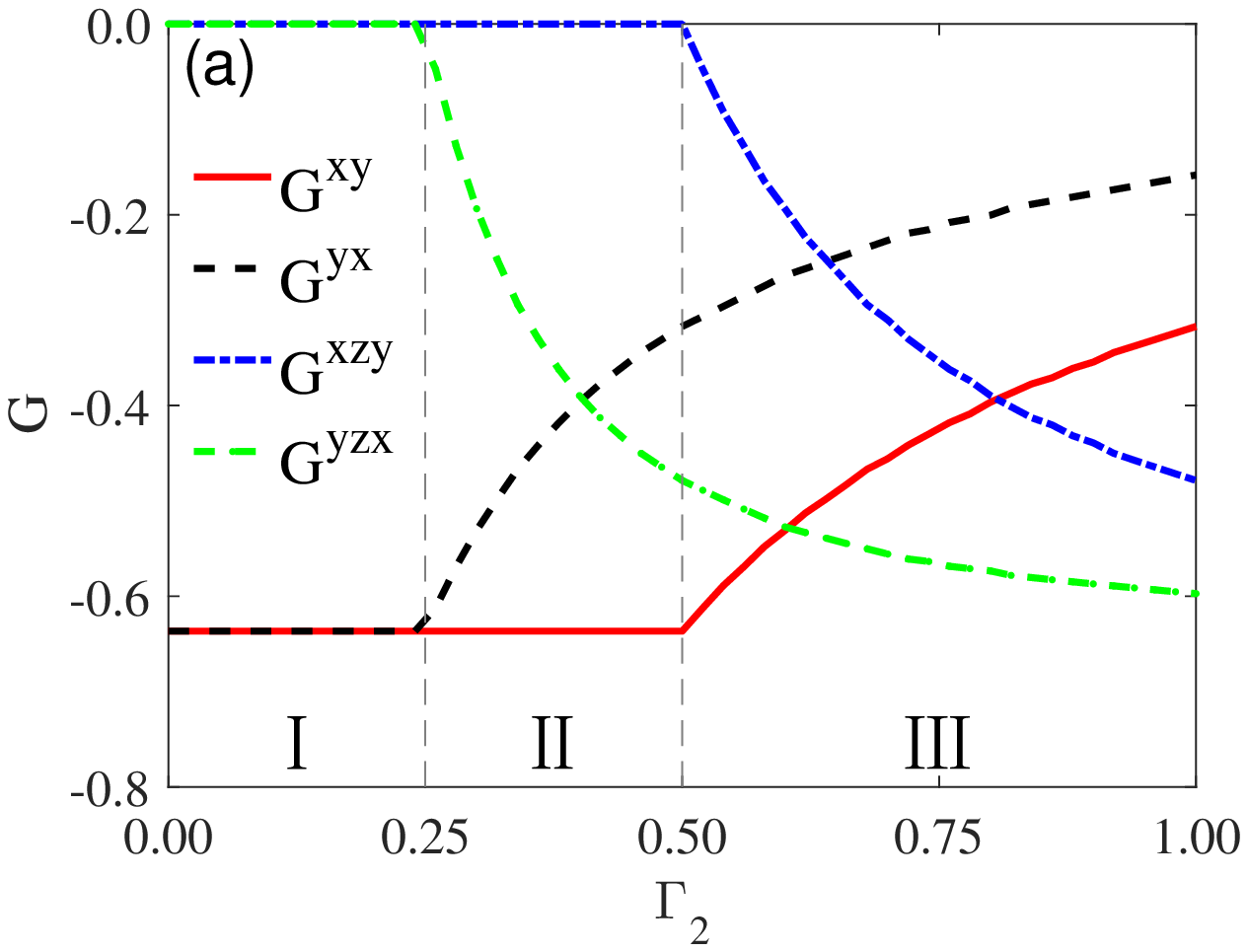}
 \includegraphics[width=7cm]{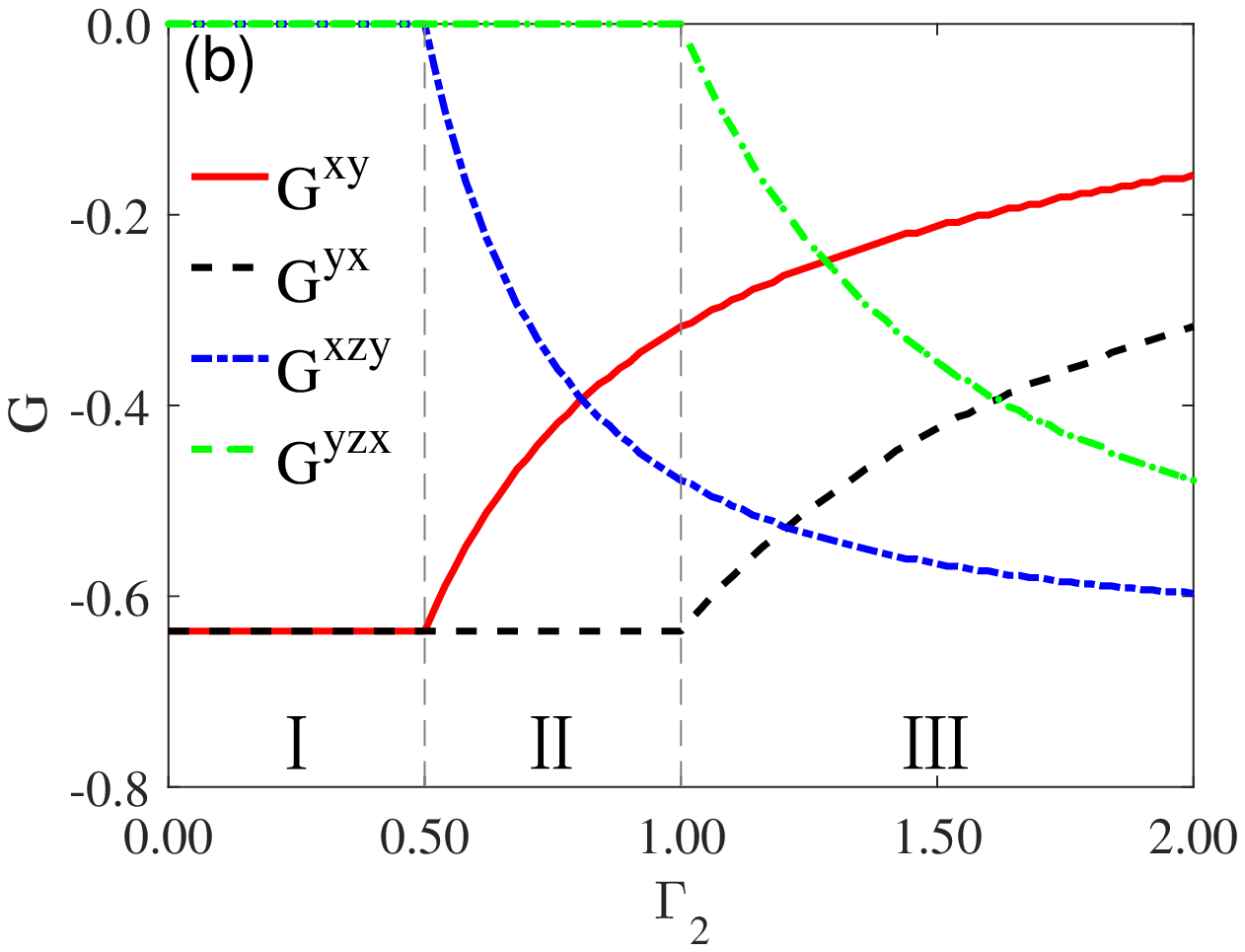}
\caption{The correlation functions $G_{j,j+1}^{xy}$, $G_{j,j+1}^{yx}$, $G_{j-1,j,j+1}^{xzy}$, $G_{j-1,j,j+1}^{yzx}$ as a function of $\Gamma_2$ for (a) $\eta=0.25$ and (b) $\eta=1$.}
\label{gl}
\end{figure}

In order to characterize the phase instability, we study the two-site correlation function $G_{i,j}^{a,b}=\langle\sigma_i^a\sigma_j^b\rangle$, where $a$, $b$=$x$, $y$, $z$. 
A simple calculation reveals that
\begin{eqnarray}
&&G_{j,j+1}^{xy}=\begin{cases}
-\frac{2}{\pi},   &   \Gamma_2\le0.5,\\
-\frac{1}{\pi\Gamma_2},  &   \Gamma_2>0.5,\\
\end{cases}
\label{Gxy}\\
&&G_{j,j+1}^{yx}=\begin{cases}
-\frac{2}{\pi},  &   \Gamma_2\le\eta,\\
-\frac{2\eta}{\pi\Gamma_2}, &   \Gamma_2>\eta.\\
\end{cases}
\label{Gyx}
\end{eqnarray}
The nearest-neighbor correlations $G_{j,j+1}^{xy}$, $G_{j,j+1}^{yx}$ show abrupt changes at $(\Gamma_{2})$$_{c,1}$=0.5 and $(\Gamma_{2})$$_{c,2}$=$\eta$. It is obvious that the QPTs at $(\Gamma_{2})$$_{c,1}$ can be attributed to Eq. (\ref{oddchain}) while the one at $(\Gamma_{2})$$_{c,2}$ originates from Eq. (\ref{evenchain}).
Similarly, the next-nearest-neighbor correlations
are defined on three consecutive spins $G_{j-1,j,j+1}^{a,b,c}$=$\langle\sigma_{j-1}^a\sigma_j^b \sigma_{j+1}^c\rangle$. We then find that
\begin{eqnarray}
 &&G_{j-1,j,j+1}^{xzy}=\begin{cases}
0,  &   \Gamma_2\le0.5,\\
\frac{1}{\pi}(\frac{1}{2\Gamma_2^2}-2),  &   \Gamma_2>0.5, \\
\end{cases}
\label{Gxzy}\\
&&G_{j-1,j,j+1}^{yzx}=\begin{cases}
0,  &   \Gamma_2\le\eta,\\
\frac{1}{\pi}(\frac{2\eta^2}{\Gamma_2^2}-2),  &   \Gamma_2>\eta. \\
\end{cases}
\end{eqnarray}
The correlation functions with respect to $\Gamma_2$ are portrayed in Fig. \ref{gl}. In phase I, the dominating nearest-neighbor correlations are $G_{j,j+1}^{xy}$=$G_{j,j+1}^{yx}$ =-$2/\pi$. Meanwhile, the next-nearest-neighbor correlations $G_{j-1,j,j+1}^{xzy}$ and $G_{j-1,j,j+1}^{yzx}$ vanish. However, the correlations become intricate in phase II, in which the dominant correlations are determined by $\eta$. For $\eta<0.5$ the magnitude of $G_{j,j+1}^{yx}$ declines and the correlation $G_{j-1,j,j+1}^{yzx}$ develops with respect to $\Gamma_2$. Analogously, for $\eta>0.5$ the correlation $G_{j-1,j,j+1}^{xzy}$ gradually prevails over $G_{j,j+1}^{xy}$ with increasing $\Gamma_2$. The growing $\Gamma_2$ eventually leads to the domination of corresponding next-nearest-neighbor correlations.
\section{Quantum entanglement and coherence}
\label{Quantum entanglement}
\begin{figure}[tb!]
\includegraphics[width=7cm]{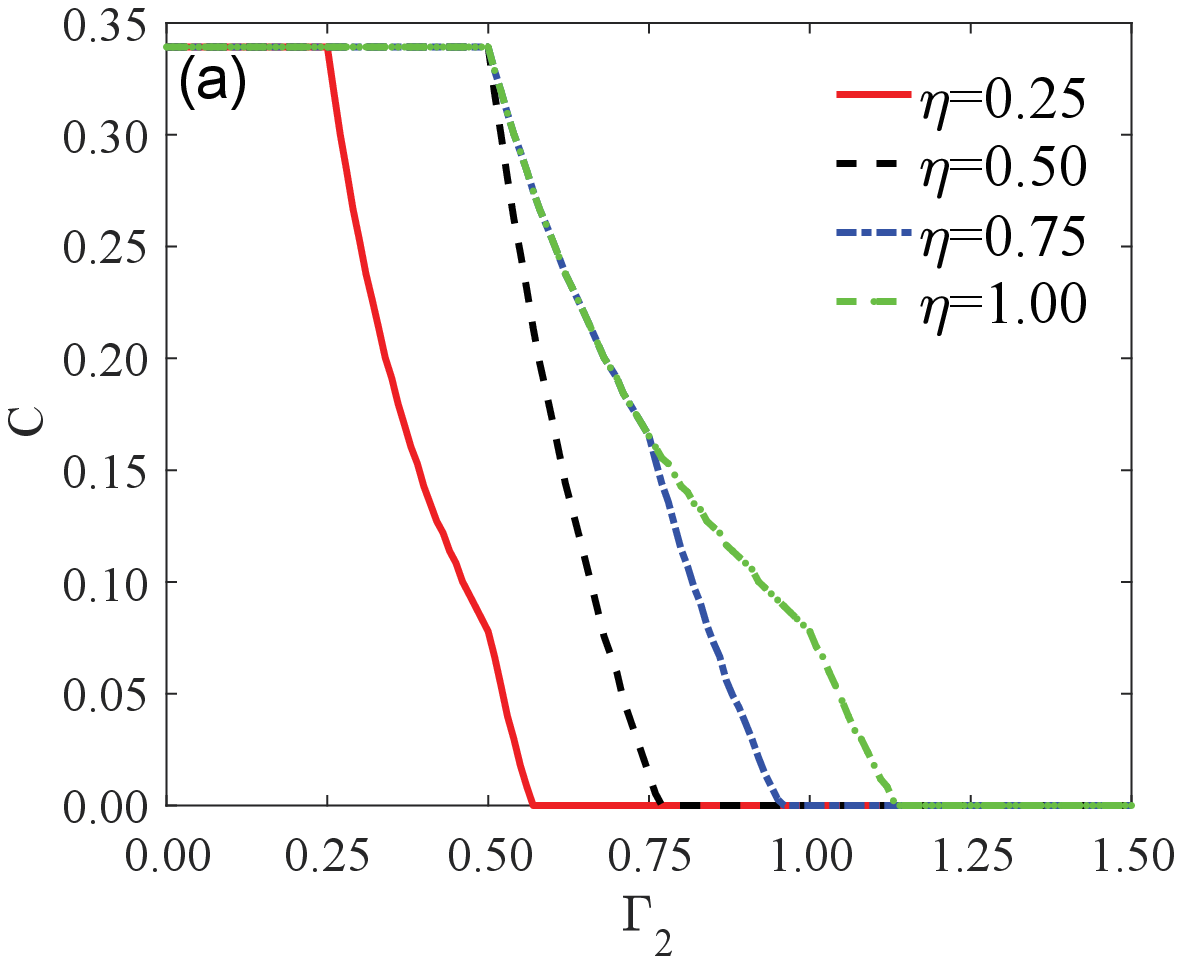}
\includegraphics[width=7cm]{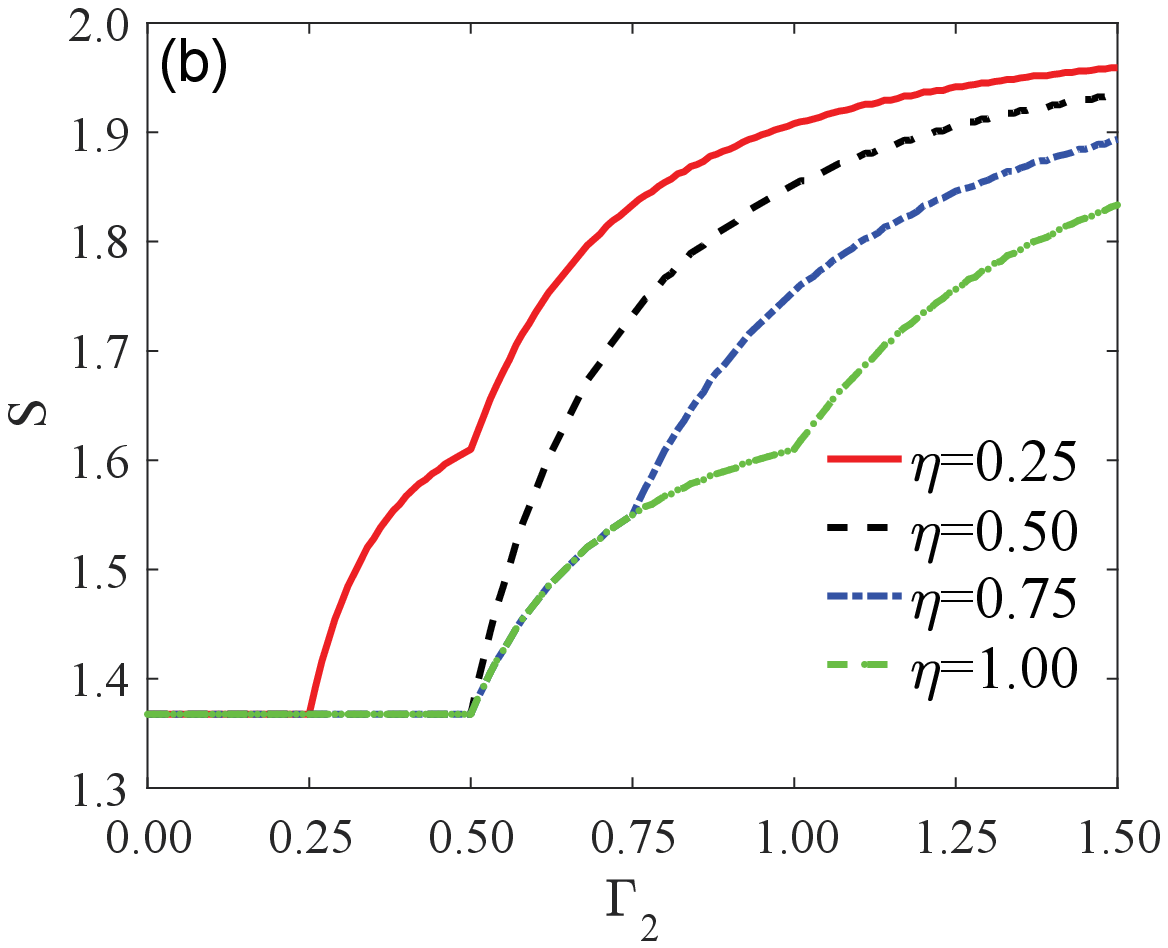}
\includegraphics[width=7cm]{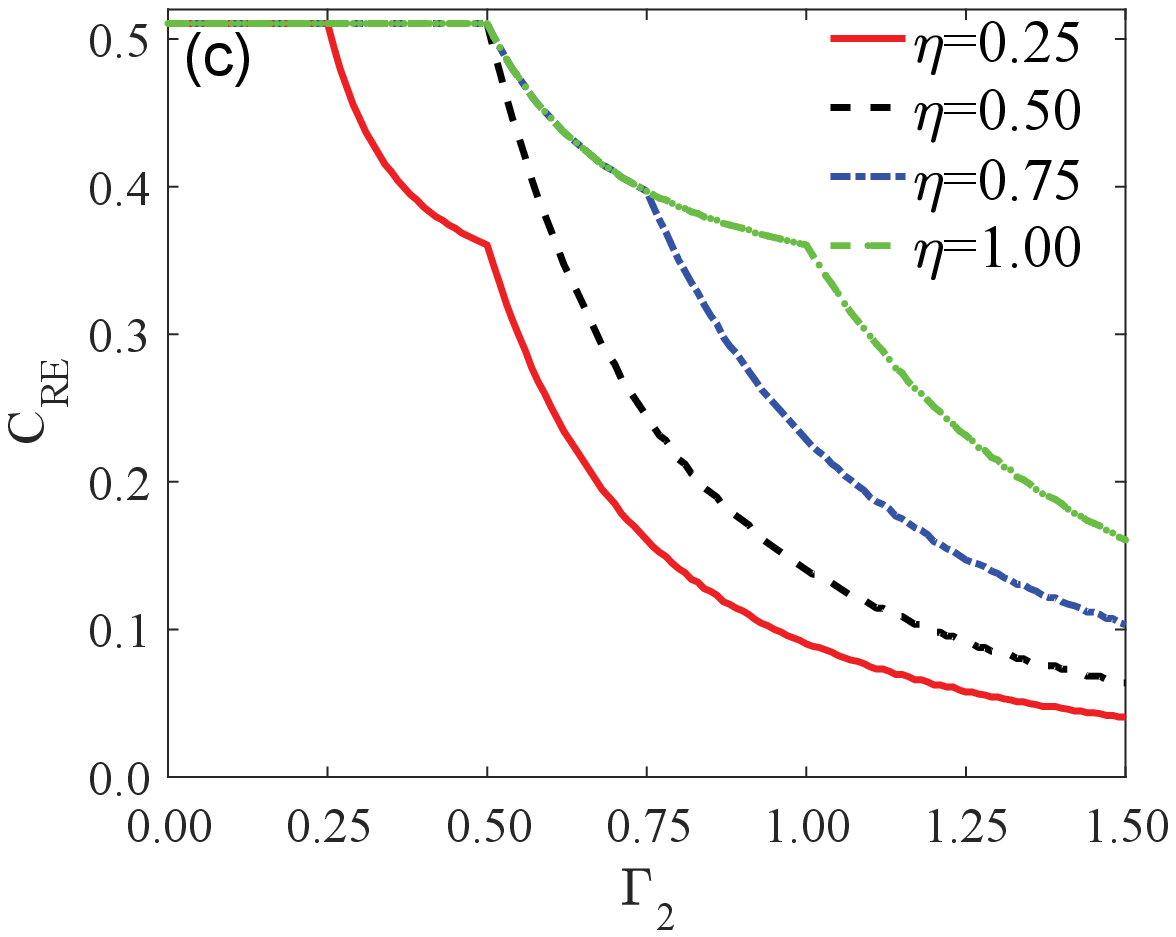}
\caption{(a) The concurrence $C_{12}$,  (b) the von Neumann entropy S
 and (c) the relative entropy $C_{RE}$ with respect to $\Gamma_2$ for $\eta=0.25,0.5,0.75,1$.}
\label{S}
\end{figure}

Since the exact solution of the Gamma model with three-site
interactions is at hand, it is straightforward
to obtain its complete information measures such as concurrence and von Neumann entropy. Concurrence is a pairwise
entanglement measure for any bipartite system that relates to
a two-site reduced density matrix $\rho_{ij}$, which is defined as:
\begin{eqnarray}
C(\rho_{ij})=\max\left\{0,r_1-r_2-r_3-r_4\right\},
\label{C}
\end{eqnarray}
where $r_n$ $(n=1, 2, 3, 4)$ are the square root of the matrix $R$ in descending order,
\begin{eqnarray}
R=\rho_{ij}(\sigma^y\otimes\sigma^y)\rho_{ij}^*(\sigma^y\otimes\sigma^y).
\end{eqnarray}
The reduced density matrix $\rho_{ij}$ of qubits $i$ and $j$ can be expressed as:
\begin{equation}
\rho_{ij}=\left(\begin{array}{cccc}
  u^+&0&0&z_1\\
   0&\omega^+&z_2&0\\
    0&z_2^*&\omega^-&0\\
     z_1^*&0&0&u^-
  \end{array}\right).
  \label{jz}
\end{equation}
with
\begin{eqnarray}
&&u^{\pm}=\frac{1}{4}(1\pm\langle\sigma_{i}^z\rangle\pm\langle\sigma_{j}^z\rangle+\langle\sigma_{i}^z\sigma_{j}^z\rangle), \\
&&z_1=\frac{1}{4}(\langle\sigma_{i}^x\sigma_{j}^x\rangle-\langle\sigma_{i}^y\sigma_{j}^y\rangle-i\langle\sigma_{i}^x\sigma_{j}^y\rangle-i\langle\sigma_{i}^y\sigma_{j}^x\rangle), \\
&&z_2=\frac{1}{4}(\langle\sigma_{i}^x\sigma_{j}^x\rangle+\langle\sigma_{i}^y\sigma_{j}^y\rangle+i\langle\sigma_{i}^x\sigma_{j}^y\rangle-i\langle\sigma_{i}^y\sigma_{j}^x\rangle), \\
&&\omega^{\pm}=\frac{1}{4}(1\pm\langle\sigma_{i}^z\rangle\mp\langle\sigma_{j}^z\rangle-\langle\sigma_{i}^z\sigma_{j}^z\rangle).
\label{ys}
\end{eqnarray}
The concurrence for such a two-qubit state $\rho_{ij}$ [Eq. (\ref{jz})] can be simplified into $C=2\max\left\{0,\varsigma_1,\varsigma_2\right\}$, where  $\varsigma_1$=$|z_1|$-$\sqrt{\omega^+\omega^-}$ and  $\varsigma_2$=$|z_2|$-$\sqrt{u^+u^-}$. To this end,
a compact form for the nearest-neighbor concurrence can be given by
\begin{eqnarray}
C_{j,j+1}=\max\left\{0,C_0\right\},
\label{Concurrence}
\end{eqnarray}
where for $\eta>0.5$,
\begin{eqnarray}
C_0=\begin{cases}
\frac{2}{\pi}(1+\frac{1}{\pi})-\frac{1}{2},  &   \Gamma_2\le0.5,\\
\frac{1}{2}(\frac{1}{\pi\Gamma_2}+\frac{2}{\pi}+\frac{2}{\pi^2\Gamma_2}-1),  &   0.5<\Gamma_2\le\eta, \\
\frac{1}{2}(\frac{1}{\pi\Gamma_2}+\frac{2\eta}{\pi\Gamma_2}+\frac{2\eta}{\pi^2\Gamma_2^2}-1),  & \Gamma_2>\eta,\\
\end{cases}
\label{C01}
\end{eqnarray}
and for $\eta<0.5$,
\begin{eqnarray}
C_0=\begin{cases}
\frac{2}{\pi}(1+\frac{1}{\pi})-\frac{1}{2},  &   \Gamma_2\le\eta,\\
\frac{1}{2}(\frac{2\eta}{\pi\Gamma_2}+\frac{2}{\pi}+\frac{4\eta}{\pi^2\Gamma_2}-1), &   \eta<\Gamma_2\le0.5, \\
\frac{1}{2}(\frac{1}{\pi\Gamma_2}+\frac{2\eta}{\pi\Gamma_2}+\frac{2\eta}{\pi^2\Gamma_2^2}-1),  & \Gamma_2>0.5.\\
\end{cases}
\label{C02}
\end{eqnarray}
The concurrence $C_{j,j+1}$ of nearest-neighbor qubits for a few sets of typical parameters is exhibited in Fig. \ref{S}(a). One finds that $C_{j,j+1}$ remains a constant in phase I and then will monotonically decrease with an increase in $\Gamma_2$ until it vanishes. We can observe the pairwise entanglement undergoes an abrupt change across the critical points.

The von Neumann entropy $S(\rho_{ij})=-{\rm Tr}(\rho_{ij}\log_2{\rho_{ij}})$ is another popular measure for bipartite correlations.  In addition, the relative entropy of coherence measure admits a distance measure between a bipartite state and its incoherent state defined as~\cite{Baumgratz14} $C_{RE}(\rho_{ij})=S(\rho_{diag})-S(\rho_{ij})$, where $S(\rho_{diag})$ represents the von Neumann entropy of the new density matrix after the nondiagonal terms are removed. Figures \ref{S}(b) and (c) show the von Neumann entropy $S$ and the relative entropy of coherence $C_{RE}$ as a function of $\Gamma_2$. One can see that they remain a constant in phase I and then present kinks at the critical points. After surpassing the first critical point, $S$ increases monotonously with increasing $\Gamma_2$,
showing an opposite trend compared to $C_{RE}$. In the large-$\Gamma_2$ limit, $C_{RE}$ eventually approaches to 0.

\section{Discussion and conclusion}\label{conclusion}
In this paper, we study the Gamma model with three-site interactions, which describes off-diagonal exchange interactions between edge-shared octahedra with strong spin-orbit couplings along a zigzag chain. The
sawtooth structure renders that the second-nearest-neighbor interactions become nonnegligible compared to the nearest-neighbor interactions. Although the Hamiltonian seems intricate, the symmetry analysis reveals it can be decoupled into two independent Majorana snake chains.
In this context, we scrutinize the ground-state properties and the associated quantum phase transitions. In terms of the Jordan-Wigner transformation, Hamiltonian (\ref{Hamiltonian1}) is transformed into a model of spinless fermions, which stabilizes a semimetallic state.
In particular, due to the frustrating nature of the second-neighbor interactions~\cite{Raghu08}, the number of Weyl nodes can be tuned through Lifshitz phase transitions. Lifshitz transitions have important applications in many areas of physics, such as high-energy physics, cosmology, black hole physics, and the search for room-$T_c$ superconductivity. For example, a black hole horizon serves as the surface of the Lifshitz transition between vacua with type-I and type-II Weyl points~\cite{Volovik18}.
We have obtained the complete phase diagram in the $\Gamma_2$-$\eta$ plane. The critical lines correspond to $(\Gamma_{2})$$_{c,1}$=$0.5$ and $(\Gamma_{2})$$_{c,2}$=$\eta$, which intersects at a multi-critical point. Although the system remains a multi-Weyl semimetals with finite density of states, there are continuous phase transitions from type-I to type-II WSMs occurring in the Brillouin zone.
The phase I is characterized by the linear dispersions at $k=0$ and $k=\pi$, which are both type-I Weyl points. As $\Gamma_2$ increases, the Weyl cones situate at $k=\pi$ are completely tilted until the Weyl node becomes type-II node in phase II, which will revert to type-I node again in phase III. It is significant
to develop realistic WSMs containing both type-I and type-II Weyl nodes simultaneously. The coexistence type-I and type-II Weyl nodes in topological materials
can motivate intriguing properties. A hybrid WSM with coexisting type-I and type-II Weyl nodes by using the tight-binding model was constructed~\cite{Li16}, and the disorder was found to induce phase transitions from type-I to type-II in WSMs~\cite{Wu17}. Recently, the material OsC$_2$ hosting 24 type-I Weyl nodes and 12 type-II Weyl
nodes was reported~\cite{Zhang18}.

Based on the exact solutions, the analytical expressions of the spin-spin correlation functions are obtained. We show that
all these measures are capable to detecting the second-order
transition. To complete the analytic approach, we present a comparative study
of diverse measures of quantum correlations including concurrence and von Neumann entropy as well as relative entropy of coherence, which undergo a sudden change in the wake of quantum phase transitions, similar to the behaviors of correlation functions.
Despite formal similarity, different measures of quantumness have their
respective trend as the strength of three-site interactions $\Gamma_2$ increases.

To summarize, we emphasize that the advantage of the
model considered here is its exact solvability that implies
in particular the possibility to calculate accurately various
dynamic quantities. Our results suggest that the Gamma model can act as a minimal model to describe Lifshitz phase transitions in correlated electron systems. The
figures of merit of this model might be crucial to understanding
off-diagonal exchange interactions with broken reflection symmetry and  provide an ideal benchmark for other computational methods and
approximate techniques used to study more realistic models.

\section*{Acknowledgments}
\label{sec_ack}
 This work was supported by the Natural Science Foundation of China (NSFC) under Grant No. 11474211 and the startup fund of Nanjing University of Aeronautics and Astronautics under Grant No. 1008-YAH20006 and stable supportfor basic institute research (Grant No. 190101).
\appendix
\section{Energy dynamics}
\label{energycurrent}

Here we show the XZY$-$YZX-type three-site interactions can emerge from the nearest-neighbor off-diagonal exchange interactions in the nonequilibrium steady states.  The Hamiltonian ${\cal H}_{\rm NN}$ describes a one-dimensional lattice with only nearest-neighbor off-diagonal exchange interactions:
 \begin{eqnarray}
{\cal H}_{\rm NN}=\sum_{l=1}^{N}[\Gamma_1(\sigma_l^x\sigma_{l+1}^y+\alpha\sigma_l^y\sigma_{l+1}^x)]=\sum_{l=1}^N \hat{h}_{l,l+1}.
  \end{eqnarray}
For simplicity, we assume that $\Gamma_1=1$. The energy current $\hat{j}_l$ in the nonequilibrium steady states is calculated by taking a time derivative
of the energy density operators and follows from the continuity
equation:
 \begin{eqnarray}
\frac{d \hat{h}_{l,l+1}}{dt}&&=i[{\cal H}_{\rm NN},\hat{h}_l]\nonumber
\\&&=-2\alpha^2\sigma_{l-1}^y\sigma_{l}^z\sigma_{l+1}^x+2\alpha^2\sigma_{l}^y\sigma_{l+1}^z\sigma_{l+2}^x\nonumber
\\&&+2\sigma_{l-1}^x\sigma_{l}^z\sigma_{l+1}^y-2\sigma_{l}^x\sigma_{l+1}^z\sigma_{l+2}^y\nonumber
\\&&=-(\hat{j}_{l+1}-\hat{j}_l)=-{\rm div} \hat{j}_l.
\end{eqnarray}
Immediately it arrives at
\begin{eqnarray}
\hat{j}_l=-2\alpha^2\sigma_{l-1}^y\sigma_{l}^z\sigma_{l+1}^x+2\sigma_{l-1}^x\sigma_{l}^z\sigma_{l+1}^y.
\end{eqnarray}
This energy current operator acts on three adjacent sites and has the $z$ component of spin-1/2 operators between two next-nearest-neighbor sites.


\begin{references}
\bibitem{Bitko96} D. Bitko, T. F. Rosenbaum, and G. Aeppli, Quantum critical behavior for a model magnet, Phys. Rev. Lett. {\bf77}, 940 (1996).

\bibitem{Coldea10} R. Coldea, D. A. Tennant, E. M. Wheeler, E. Wawrzynska, D. Prabhakaran, M. Telling, K. Habicht, P. Smeibidl, and K. Kiefer, Quantum criticality in an Ising chain: Experimental
evidence for emergent E$_8$ symmetry, Science {\bf327}, 177 (2010).

\bibitem{Kenzelmann02} M. Kenzelmann, R. Coldea, D. A. Tennant, D. Visser, M. Hofmann, P. Smeibidl, and Z. Tylczynski, Order-to-disorder transition in the XY-like quantum magnet Cs$_2$CoCl$_4$ induced by
noncommuting applied fields, Phys. Rev. B {\bf65}, 144432 (2002).

\bibitem{Breunig13} O. Breunig, M. Garst, E. Sela, B. Buldmann, P. Becker, L. Bohat\'{y}, R. M\"{u}ller, and T. Lorenz, Spin-1/2 XXZ chain system Cs$ _2$CoCl$_4$ in a transverse magnetic field, Phys. Rev. Lett. {\bf111}, 187202 (2013).

\bibitem{Breunig17} O. Breunig, M. Garst, A. Kl\"{u}mper, J. Rohrkamp, M. M. Turnbull, and T. Lorenz, Quantum criticality in the spin-1/2 Heisenberg chain system copper pyrazine dinitrate, Sci. Adv. {\bf 3}, 12 (2017).

\bibitem{Kit06} A. Kitaev, Anyons in an exactly solved model and beyond,
                   Ann. Phy. \textbf{321}, 2 (2006).

\bibitem{Jackeli09} G. Jackeli and G. Khaliullin, Mott Insulators in the Strong Spin-Orbit Coupling Limit: From Heisenberg to a Quantum Compass and Kitaev Models, Phys. Rev. Lett. {\bf 102}, 017205 (2009).

\bibitem{Chaurasiya19} A. k.  Chaurasiya, A. Kumar, R. Gupta, S. Chaudhary, P. K. Muduli, and A. Barman, Direct observation of unusual interfacial Dzyaloshinskii-Moriya interaction in graphene/NiFe/Ta heterostructures, Phys. Rev. B {\bf99}, 035402 (2019).

\bibitem{You2014}
{W.-L. You, P. Horsch, and A. M. Ole\'s, Quantum phase transitions in exactly solvable one-dimensional compass models,
Phys. Rev. B {\bf89}, 104425 (2014).}

\bibitem{Takikawa19} D. Takikawa and S. Fujimoto, Impact of off-diagonal exchange interactions on the Kitaev spin-liquid state of $\alpha$-RuCl$_3$, Phys. Rev. B {\bf 99}, 224409 (2019).

\bibitem{Rau2014}
 J. G. Rau, E. K.-H. Lee, and H.-Y. Kee, Generic Spin Model for the Honeycomb Iridates beyond the Kitaev Limit, Phys. Rev. Lett. {\bf 112},
 077204 (2014).

\bibitem{Gordon2019}Jacob S. Gordon, Andrei Catuneanu, Erik S. S\o{}rensen and Hae-Young Kee,  Theory of the field-revealed Kitaev spin liquid, Nat. Commun. {\bf 10}, 2470 (2019).

\bibitem{Yang20}Wang Yang, Alberto Nocera, Tarun Tummuru, Hae-Young Kee, and Ian Affleck, Phase Diagram of the Spin-1/2 Kitaev-Gamma Chain and Emergent SU(2) Symmetry, Phys. Rev. Lett. {\bf 124}, 147205 (2020).

\bibitem{Yang202}Wang Yang, Alberto Nocera, Erik S. S\o{}rensen, Hae-Young Kee, and Ian Affleck, Spin-nematic order in the spin-1/2 Kitaev-Gamma chain, arXiv:2004.06074.


\bibitem{Gardner10} J. S. Gardner, M. J. P. Gingras, and J. E. Greedan, Magnetic pyrochlore oxides, Rev. Mod. Phys. {\bf82}, 53 (2010).



\bibitem{Rynbach10} A. van Rynbach, S. Todo, and S. Trebst, Orbital Ordering in $e_g$ Orbital Systems: Ground States and Thermodynamics of the 120$^{\circ}$ Model, Phys. Rev. Lett.
{\bf105}, 146402 (2010).

\bibitem{Dzyaloshinsky1958}
{I. Dzyaloshinsky, A thermodynamic theory of "weak" ferromagnetism of antiferromagnetics, J. Phys. Chem. Solids {\bf4}, 241 (1958).}

\bibitem{Moriya1960}
{T. Moriya, Anisotropic Superexchange Interaction and Weak Ferromagnetism, Phys. Rev. {\bf 120}, 91 (1960).}

\bibitem{yi2019}
{T.-C. Yi, W.-L. You, N. Wu, and A. M. Ole\'s, Criticality and factorization in the Heisenberg chain with Dzyaloshinskii-Moriya interaction, Phys. Rev. B {\bf100}, 024423 (2019).}

\bibitem{SatyakiKar2018}
{S. Kar and B. Basu, Photoinduced entanglement in a magnonic Floquet topological insulator, Phys. Rev. B {\bf98}, 245119 (2018).}

\bibitem{You142}W.-L. You, G.-H. Liu, P. Horsch, and A. M. Ole\'s,  Exact treatment of magnetism-driven ferroelectricity in the one-dimensional compass model, Phys. Rev. B {\bf 90}, 094413 (2014).

\bibitem{Qiu16}Y.-C. Qiu, Q.-Q. Wu and W.-L. You,  Energy dynamics in a generalized compass chain,  J. Phys.: Condens. Matter {\bf 28}, 496001 (2016).

\bibitem{WuQC17}  Q.-Q. Wu, W.-H. Ni and W.-L. You, J. Phys.: Condens. Matter {\bf 29}, 225804 (2017).

\bibitem{Huia2017}
{N. Hui, Y. Xu, J. Wang, Y. Zhang, and Z. Hu, Quantum coherence and quantum phase transition in the XY model with staggered Dzyaloshinsky-Moriya interaction, Physica B {\bf 510}, 7 (2017) .}

\bibitem{Liang2017}
{L. Qiu, D. Quan, F. Pan, and Z. Liu, Skew information in the XY model with staggered Dzyaloshinskii-Moriya interaction, Physica B {\bf 514}, 13 (2017).}

\bibitem{Ellen2019}
{E. Fogh, O. Zaharko, J. Schefer, C. Niedermayer, S. Holm-Dahlin, M. K. S\o rensen, A. B. Kristensen, N. H. Andersen, D. Vaknin, N. B. Christensen, and R. Toft-Petersen, Dzyaloshinskii-Moriya interaction and the magnetic ground state in magnetoelectric LiCoPO$_4$, Phys. Rev. B {\bf99}, 104421 (2019).}

\bibitem{Kenzelmann2006}
 {M. Kenzelmann, A. B. Harris, A. Aharony, O. Entin-Wohlman,
T. Yildirim, Q. Huang, S. Park, G. Lawes, C. Broholm,
N. Rogado, R. J. Cava, K. H. Kim, G. Jorge, and A. P. Ramirez, Field dependence of magnetic ordering in kagom\'e-staircase compound Ni$_3$V$_2$O$_8$, Phys. Rev. B {\bf74}, 014429 (2006).}

\bibitem{Dhital2017}
{C. Dhital, L. DeBeer-Schmitt, Q. Zhang, W. Xie, D. P. Young, and J. F. DiTusa, Exploring the origins of the Dzyaloshinskii-Moriya interaction in MnSi,
Phys. Rev. B {\bf96}, 214425 (2017).}

\bibitem{Shanavas2016}
{K. V. Shanavas and S. Satpathy, Electronic structure and the origin of the Dzyaloshinskii-Moriya interaction in MnSi,
Phys. Rev. B {\bf93}, 195101 (2016).}

\bibitem{Soucaille2016}
{R. Soucaille, M. Belmeguenai, J. Torrejon, J.-V. Kim, T. Devolder, Y. Roussign\'e, S.-M. Ch\'erif, A. A. Stashkevich, M. Hayashi, and J.-P. Adam, Probing the Dzyaloshinskii-Moriya interaction in CoFeB ultrathin films using domain wall creep and Brillouin light spectroscopy,
Phys. Rev. B {\bf94}, 104431 (2016).}

\bibitem{Zotos1997}
{X. Zotos, F. Naef, and P. Prelov\v sek, Transport and conservation laws, Phys. Rev. B {\bf55}, 11029 (1997).}

\bibitem{Antal1997}
{T. Antal, Z. R\'acz, and L. Sas\'vari, Nonequilibrium Steady State in a Quantum System: One-Dimensional Transverse Ising Model with Energy Current, Phys. Rev. Lett. {\bf78}, 167 (1997).}

\bibitem{Tong2015}
{S. Lei and P. Tong, Quantum discord in the transverse field XY chains with three-spin interaction, Physica B (Amsterdam) {\bf463}, 1 (2015).}

\bibitem{Tong2016}
{S. Lei and P. Tong, Wigner-Yanase skew information and quantum phase transition in one-dimensional quantum spin-1/2 chains, Quantum Inf. Process. {\bf 15}, 1811 (2016).}

\bibitem{You2016}
W.-L. You, Y.-C. Qiu and A. M. Ole\'s, Quantum phase transitions in a generalized compass chain with three-site interactions, Phys. Rev. B {\bf 93}, 214417 (2016).

\bibitem{Lieb61}
E. Lieb, T. Schultz and D. Mattis, Two soluble models of an antiferromagnetic chain,  Ann. Phys.  {\bf 16}, 407 (1961).

\bibitem{Cabrera87}G. G. Cabrera and R. Jullien, Role of boundary conditions in the finite-size Ising model, Phys. Rev. B {\bf 35}, 7062 (1987).

\bibitem{Wu20}
N. Wu, Longitudinal magnetization dynamics in the quantum Ising ring: A Pfaffian method
based on correspondence between momentum space and real space, Phys. Rev. E {\bf 101,}, 042108 (2020).

\bibitem{You08}
W.-L. You and G.-S. Tian, Quantum phase transition in the one-dimensional compass model using the pseudospin approach. Phys. Rev. B {\bf 78}, 184406 (2008).
\bibitem{You14}
W.-L. You, P. Horsch, and A. M. Ole\'s, Quantum phase transitions in exactly solvable one-dimensional compass models, Phys. Rev. B
{\bf 89}, 104425 (2014).

\bibitem{Nussinov14}
Z. Nussinov and J. van den Brink, Compass models: Theory and physical motivations, Rev. Mod. Phys. {\bf 87}, 1 (2015).

\bibitem{Wu19}
N. Wu and W.-L. You, Exact zero modes in a quantum compass chain under inhomogeneous transverse fields, Phys. Rev. B {\bf 100}, 085130 (2019).

\bibitem{Baumgratz14}
T. Baumgratz, M. Cramer, and M. B. Plenio, Quantifying Coherence, Phys. Rev. Lett. {\bf 113}, 140401 (2014).

\bibitem{Raghu08}
S. Raghu, X.-L. Qi, C. Honerkamp, and S.-C. Zhang, Topological Mott Insulators, Phys. Rev. Lett. {\bf 100}, 156401 (2008).

\bibitem{Volovik18}
G E Volovik, Exotic Lifshitz transitions in topological materials,
Phys.-Usp. {\bf 61}, 89 (2018).

\bibitem{Li16}
{F. Li, X. Luo, X. Dai, Y. Yu, F. Zhang, and G. Chen, Hybrid Weyl semimetal, Phys. Rev. B {\bf 94}, 121105(R)  (2016).}

\bibitem{Wu17}
{Y. Wu, H. Liu, H. Jiang, and X. C. Xie, Global phase diagram of disordered type-II Weyl semimetals, Phys. Rev. B {\bf96}, 024201 (2017)}.

\bibitem{Zhang18}
{M. Zhang, Z. C. Yang, and G. Wang, Coexistence of Type-I and Type-II Weyl Points in the Weyl-Semimetal OsC$_2$, J. Phys. Chem. C {\bf122}, 3533 (2018).}

\end{references}
\end{document}